\documentclass[11pt]{article}
\usepackage{moriond,epsfig}
\usepackage{amsmath}
\usepackage{amssymb}
\usepackage{wasysym}
\usepackage{graphicx}

\def\Journal#1#2#3#4{{#1} {\bf #2}, #3 (#4)}

\def\NPB{{\em Nucl. Phys.} B}
\def\PLB{{\em Phys. Lett.}  B}
\def\PRL{\em Phys. Rev. Lett.}
\def\PRD{{\em Phys. Rev.} D}

\def\be{\begin{equation}}
\def\ee{\end{equation}}
\def\bea{\begin{eqnarray}}
\def\eea{\end{eqnarray}}

\newcommand{\op}{{\cal Q}}
\newcommand{\nn}{\nonumber}
\newcommand{\ii}{\mathrm{i}}

\begin{document}
\vspace*{4cm}
\title{Top, Polarization, LHC and New Physics}

\author{ J.F. KAMENIK }

\address{J. Stefan Institute, Jamova 39, P. O. Box 3000,\\ 
1001 Ljubljana, Slovenia}
\address{Department of Physics, University of Ljubljana, Jadranska 19, \\
1000 Ljubljana, Slovenia}

\maketitle\abstracts{
Polarization observables in top quark decays are sensitive probes of possible new physics contributions to the interactions of the heavy third generation quarks.  Within an effective theory approach such new physics contributions can be classified in terms of several higher dimensional operators. We investigate the interplay between indirect constraints on such operators, coming mainly from rare B physics processes, and direct measurements of top polarization observables at the LHC.}

\section{Introduction}
The extensive production of top quarks at the LHC and Tevatron colliders offers the possibility to study $t W b$ interactions with high accuracy. 
Within the Standard Model (SM) the partial $t \to b W$ decay width and the branching fraction
\be
\Gamma(t\to b W)^{\rm SM} \simeq \frac{\alpha |V_{tb}|^2}{16 s_W^2} \frac{m_t^3}{m_W^2}\,, \qquad \mathcal B(t\to b W)^{\rm SM} \simeq \frac{|V_{tb}^2|}{|V_{tb}^2|+|V_{ts}^2|+|V_{td}^2|}\,,
\ee
are sensitive to the value of the CKM matrix element $V_{tb}$, related to the top-bottom charged current.\cite{Alwall:2006bx} However, present indirect constraints on $V_{tb}$ within the SM employing CKM unitarity\cite{PDG} are already much stronger compared to the present\,\cite{Group:2009qk} and projected\,\cite{LHC} experimental direct sensitivity.

Fortunately, helicity fractions of the final state $W$ in this decay provide additional information on the structure of the $tWb$ interaction. Considering leptonically decaying $W$'s, one can define the angle between the charged lepton momentum in the $W$ rest frame and the $W$ momentum in the $t$-quark rest frame ($\theta_\ell^*$). Then the normalized differential decay rate for unpolarized top quarks can be written as 
\be
\frac{1}{\Gamma} \frac{d\Gamma}{d\cos\theta_\ell^*} = \frac{3}{8}(1+\cos\theta_\ell^*)^2 \mathcal F_+ + \frac{3}{8} (1-\cos\theta_\ell^*)^2 \mathcal F_- + \frac{3}{4} \sin^2\theta_\ell^* \mathcal F_L\,,
\ee
with $\mathcal F_i = \Gamma_i / \Gamma$ being the $W$-boson helicity fractions.\cite{Castro2008,Dalitz1992} \footnote{Note that by definition $\sum_i \mathcal F_i = 1$ so that only two of the helicity fractions represent independent observables\,.}
There has been a continuing interest in the measurement of $\mathcal F_i$ by the CDF and D\O \, collaborations at the Tevatron. Their most recent analyses yield~\cite{Aaltonen:2010ha,Abazov:2010jn}
\begin{align}
\mathcal F^{\rm CDF}_L &= 0.88(13)\,, & \mathcal F_L^{\rm D\O} &= 0.669(102)\,, \nonumber\\
\mathcal F^{\rm CDF}_+ &= -0.15(9)\,, & \mathcal F_+^{\rm D\O} &= 0.023(53)\,,
\label{eq:1}
\end{align}
where the statistical and systematic uncertainties have been combined in quadrature. Compared to these values, an order of magnitude improvement in precision  is expected from the LHC experiments in the coming years.~\cite{Castro2008,AguilarSaavedra:2007rs}

In the SM, simple helicity considerations show that $\mathcal F_+$ vanishes at the Born level in the $m_b = 0$ limit. A non-vanishing $\mathcal F_+$ could arise from i) $m_b \neq 0$ effects, ii) $\mathcal O (\alpha_s)$ radiative corrections due to gluon emission~\footnote{Electroweak corrections also contribute, but turn out to be much smaller.~\cite{Do:2002ky}}, or from iii) non-SM $tWb$ interactions. The $\mathcal O (\alpha_s)$ and the $m_b\neq 0$ corrections to $\mathcal F_+$ have been shown to occur only at the per-mille level in the SM.~\cite{Fischer:2000kx} Specifically, they yield
\be
\mathcal F_L^{\rm SM} = 0.687(5)\,, \qquad \mathcal F_+^{\rm SM} = 0.0017(1)\,.
\ee
One could therefore conclude that measured values of $\mathcal F_+$ exceeding 0.2\% level, would signal the presence of new physics (NP) beyond the SM.

\section{Effective theory analysis}

The structure of NP contributions possibly affecting $t\to b W$ transitions can be analyzed using effective field theory methods -- by introducing the effective Lagrangian
\begin{eqnarray}
{\cal L}={\cal L}_{\mathrm{SM}}+\frac{1}{\Lambda^2}\sum_i C_i \mathcal Q_i +\mathrm{h.c.}+ {\cal O}(1/\Lambda^3)\,,
\label{eq:lagr}
\end{eqnarray}
where ${\cal L}_{\mathrm{SM}}$ is the SM part, $\Lambda$ is the scale of NP and $\mathcal Q_i$ are dimension-six operators, invariant under SM gauge transformations and consisting of SM fields.  In order to exhibit observable effects in the $t\to b W$ decays $\mathcal Q_i$ should also not mediate flavor changing neutral currents (FCNCs) in the down sector at the tree-level.\cite{Grzadkowski:2008mf}
Since the SM electroweak symmetry breaking induces misalignment between the up and down quark mass eigenbases via the CKM mechanism, isolating NP effects in $tWd_j$ interactions to a particular single flavor transition in the physical (mass) basis in general requires a large degree of fine-tuning in the flavor structure of the effective operators at the high scale, where they are generated. One possible solution is to require the operators to be flavor aligned with either the up or the down Yukawas of the SM resulting effectively in minimal flavor violating (MFV) scenarios.\cite{MFV} A systematic analysis of all MFV allowed flavor structures even in the presence of large bottom Yukawa effects yields a total of seven dimension-six effective operators which can significantly affect the $tWb$ interaction~\cite{Drobnak:2011wj}
\begin{eqnarray}
\nn \mathcal Q_{LL}&=&[\bar Q^{\prime}_3\tau^a\gamma^{\mu}Q'_3] \big(\phi_d^\dagger\tau^a\ii D_{\mu}\phi_d\big) \hspace{-0.1cm}-\hspace{-0.1cm}[\bar Q'_3\gamma^{\mu}Q'_3]\big(\phi_d^\dagger\ii D_{\mu}\phi_d\big),\\
\nn \mathcal Q_{LRt} &=& [\bar Q'_3 \sigma^{\mu\nu}\tau^a t_R]{\phi_u}W_{\mu\nu}^a \,,\\
 \mathcal Q_{RR}&=& V_{tb} [\bar{t}_R\gamma^{\mu}b_R] \big(\phi_u^\dagger\ii D_{\mu}\phi_d\big) \,, \nn\\
 \mathcal Q_{LRb} &=& [\bar Q_3 \sigma^{\mu\nu}\tau^a b_R]\phi_d W_{\mu\nu}^a \,, \nn\\
\nn \mathcal Q'_{LL}&=&[\bar Q_3\tau^a\gamma^{\mu}Q_3] \big(\phi_d^\dagger\tau^a\ii D_{\mu}\phi_d\big) \hspace{-0.1cm}-\hspace{-0.1cm}[\bar Q_3\gamma^{\mu}Q_3]\big(\phi_d^\dagger\ii D_{\mu}\phi_d\big),\\
\nn \mathcal Q^{\prime\prime}_{LL}&=&[\bar Q'_3\tau^a\gamma^{\mu}Q_3] \big(\phi_d^\dagger\tau^a\ii D_{\mu}\phi_d\big) \hspace{-0.1cm}-\hspace{-0.1cm}[\bar Q'_3\gamma^{\mu}Q_3]\big(\phi_d^\dagger\ii D_{\mu}\phi_d\big),\\
 \mathcal Q'_{LRt} &=& [\bar Q_3 \sigma^{\mu\nu}\tau^a t_R]{\phi_u}W_{\mu\nu}^a \,,\label{operators}
\label{eq:ops2}
\end{eqnarray}
where we have introduced $Q_3=(V^*_{kb} u_{Lk},b_{L})$, $Q'_3  = ({t}_L,V_{ti} {d}_{iL})$, $\sigma^{\mu\nu}=\ii [\gamma^{\mu},\gamma^{\nu}]/2$ and  $W^a_{\mu\nu}=\partial_{\mu}W_{\nu}^a-\partial_{\nu}W_{\mu}^a - g\epsilon_{abc}W_{\mu}^b W_{\nu}^c$. Furthermore, $q_{L(R)}=P_{L(R)}q$ denote the left- and right-handed quark fields ($q=u_i,d_i$), where $P_{L(R)} = (1\mp\gamma_5)/2$, while $\phi_{u,d}$ are the up- and down-type Higgs fields (in the SM $\phi_u =\ii \tau^2 \phi_d^*$) and $g$ is the weak coupling constant. The first two operators in (\ref{operators}) appear already at zeroth order in the down-type Yukawa insertions, the following two would be linear in a bottom Yukawa expansion, while the remaining three necessarily require the insertion of at least two down-type Yukawa matrices.

In the mass basis all of these operators contribute to the four possible helicity structures of the $tWb$ vertex
\be
\mathcal O_{L(R)} = \frac{g}{\sqrt 2} W^-_\mu \left[ \bar b \gamma^\mu P_{L(R)} t \right] \,, \qquad \mathcal O_{LR(RL)} = \frac{g}{\sqrt 2} W^-_{\mu\nu} \left[ \bar b \sigma^{\mu\nu} P_{L(R)} t \right] \,.
\label{eq:O}
\ee
However, at the same time they also enter FCNCs in the $B$ meson sector at one-loop resulting in severe constraints from $B\to X_s\gamma$ branching ratio measurements\,\cite{Grzadkowski:2008mf} and $B_{s,d}$ meson oscillation observables.\cite{Drobnak:2011wj} Presently, the operator $\mathcal O_{LR}$ is least affected by these indirect constraints and thus has the potential to modify the $t\to bW$ decay characteristics in an observable way. However its contributions to $\mathcal F_+$ exhibit the same helicity suppression as the SM, mandating the evaluation of $t\to bW$ decay in presence of such NP contributions at next-to-leading order (NLO) in QCD.\cite{Drobnak:2010ej} After taking into account the existing indirect bounds on the operators in (\ref{eq:O}), contributions of $\mathcal O_{LR}$ indeed allow for largest enhancement in $\mathcal F_+$, but are still necessarily below $2\permil$. Turning to $\mathcal F_L$, observable effects due to most operators in (\ref{eq:O}) are again suppressed due to indirect constraints from $B$ FCNCs, with the exception of $\mathcal O_{LR}$  where direct measurements at the Tevatron~\cite{Aaltonen:2010ha,Abazov:2010jn} are already providing competitive constraints (see figure~\ref{fig:1}).

\section{Interplay with new CP violating contributions in $B_{d,s}-\bar B_{d,s}$ oscillations}

Recently, possible NP effects in the $B_{d,s}-\bar B_{d,s}$, mixing amplitudes have received considerable attention. In particular within the SM, the $B^0-\bar B^0$ mass difference and the time-dependent CP asymmetry in $B\to J/\psi K_s$ are strongly correlated with the branching ratio $\mathrm{Br}(B^+\to \tau^+ \nu)$.\cite{Deschamps:2008de} The most recent global analyses point to a disagreement of this correlation with direct measurements at the level of 2.9 standard deviations.\cite{Lenz:2010gu} Similarly in the $B_s$ sector the recently measured CP-asymmetries by the Tevatron experiments, namely in $B_s \to J/\psi\phi$~\cite{betas} and in di-muonic inclusive decays~\cite{Abazov:2010hj} when combined, deviate from the SM prediction for the CP violating phase in $B_s-\bar B_s$ mixing by 3.3 standard deviations.\cite{Lenz:2010gu} 

\begin{figure}
\begin{center}
\includegraphics[scale= 0.4]{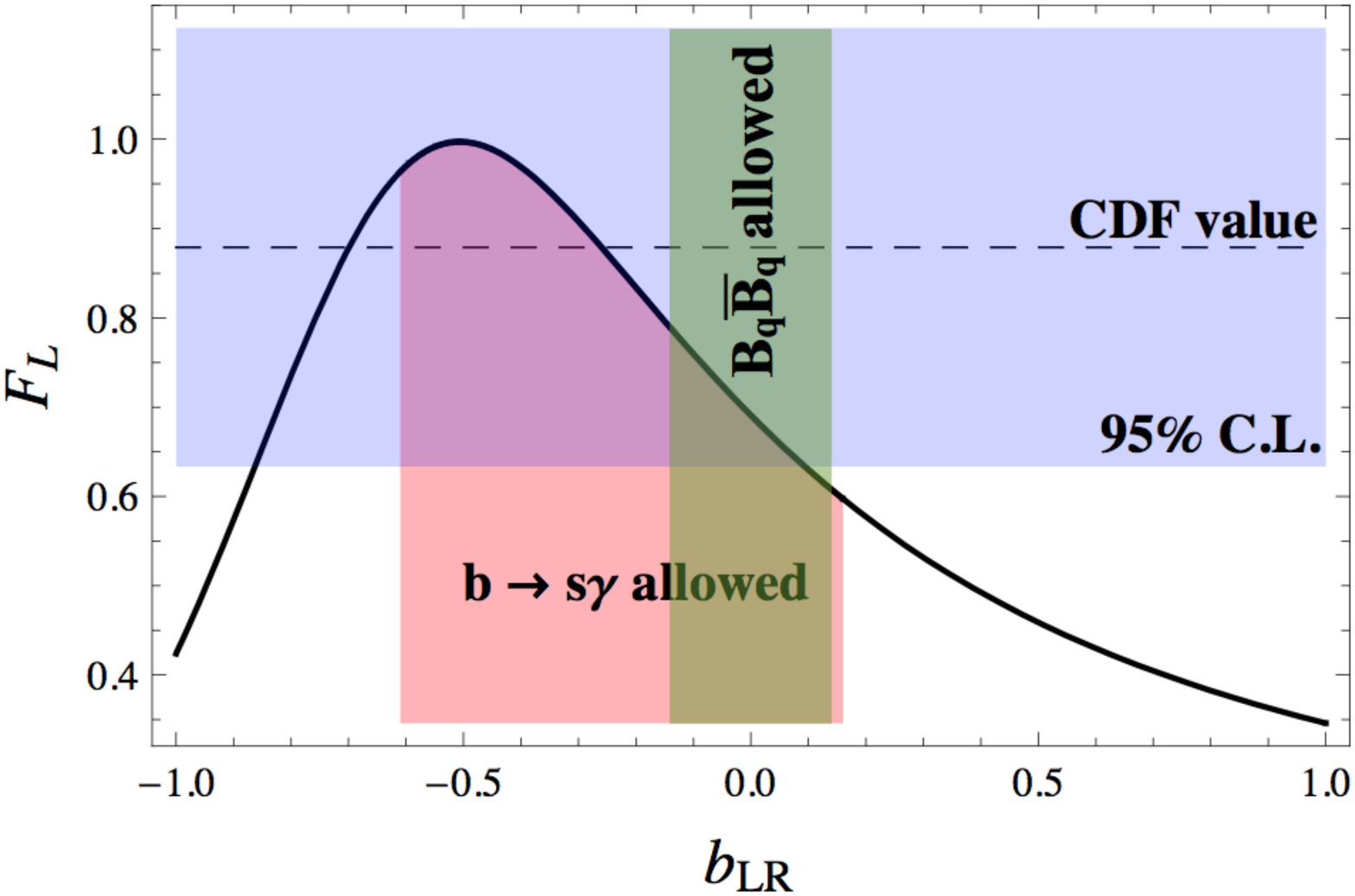}
\end{center}
\caption{
Prediction of $\mathcal F_L$ as a function of the normalized Wilson coefficient $b_{LR}=C_{LRt} v m_t / \Lambda^2$, where $v=246$~GeV corresponding to the effective operator $\mathcal Q_{LRt}$ in (\ref{operators}) (other possible NP contributions being set to zero). 
Red band below the full curve shows the allowed interval for $b_{LR}$ as given by the $B\to X_s\gamma$ analysis\,
while the vertical green band denotes values allowed $B_{s,d}$ oscillation data, both at $95\%$ C.L.\,.
For comparison, we also show the recent CDF measurement of $\mathcal F_L$ given in eq. (\ref{eq:1}) (horizontal blue shaded region).
\label{fig:1}}
\end{figure}

Anomalous $tWd_j$ interactions offer a possible solution of these anomalies via their contributions to $B_{d,s}-\bar B_{d,s}$ oscillation observables at the one-loop level. Within the MFV approach they contribute universally to $B_d$ and $B_s$ mixing amplitudes.\cite{Drobnak:2011wj} Such case has been analyzed in general~\cite{Lenz:2010gu,Ligeti:2010ia} and found consistent with present data. Among the operators in ({\ref{operators}}), contributions of $\mathcal Q_{RR}$ and $\mathcal Q_{LRb}$ to $B_{s,d}$ oscillations are severely suppressed by constraints coming from the $B\to X_s\gamma$ decay.\cite{Grzadkowski:2008mf} On the other hand, contributions of operators  $\mathcal Q_{LL}$ and $\mathcal Q_{LRt}$ cannot introduce new CP violating phases. Namely as shown recently,\cite{Blum:2009sk} a necessary condition for new flavor violating structures $\mathcal Y_x$ to introduce new sources of CP violation in quark transitions  is that ${\rm Tr}(\mathcal Y_x[Y_u Y_u^\dagger  ,  Y_d Y_d^\dagger ])\neq 0$, where $Y_{u,d}$ are the SM up- and down-quark Yukawa matrices. In MFV models (where $\mathcal Y_x$ is built out of $Y_u$ and $Y_d$ ) this condition can only be met if $\mathcal Y_x$ contains products of both $Y_u$ and $Y_d$. In (\ref{operators}) this is true for all operators except $\mathcal Q_{LL}$ and $\mathcal Q_{LRt}$. One can still use the present oscillation data to put bounds on contributions of these two operators. In particular, the resulting indirect constraint on the $\mathcal O_{LR}$ structure contributing to $t\to b W$ decay is comparable to both, the indirect $B\to X_s\gamma$ bound due to the same $\mathcal Q_{LRt}$ operator, as well the present direct $\mathcal F_L$ measurements as shown in figure~\ref{fig:1}.
The remaining three operators in (\ref{operators}), $\op_{LL}^{\prime(\prime\prime)}$ and $\op'_{LRt}$ can contribute with new CP violating phases and are not overly constrained by the $B \to X_s \gamma$ decay rate measurement.\cite{Drobnak:2011wj} As such they can account for the recently observed anomalies in the CP violating observables related to $B_{s,d} - \bar B_{s,d}$ mixing. 

Finally, one can try to predict the effects of effective operators in (\ref{operators}) on the helicity fractions of the $W$ boson in the $t\to b W$ decay channel, provided these same operators are responsible for new CP violating contributions in $B_{d,s}$ meson mixing. Both $\op_{LL}^{\prime(\prime\prime)}$ have the same chiral structure as the SM contribution and thus cannot affect the helicity fractions. They only yield small corrections to the total $t\to b W$ decay rate. On the other hand $\op'_{LRt}$ contributes to the helicity structure $\mathcal O_{LR}$. Under its influence, ${\cal F}_{L,+}$ can deviate by as much as $15\%$ and $30\%$ respectively compared to the SM predictions, although much smaller deviations are perfectly consistent with the ranges for the relevant Wilson coefficient of $\op'_{LRt}$ preferred by the $B_{d,s}$ mixing analysis. A robust prediction that can be made however is that at least one of the two independent helicity fractions (${\cal F}_{L,+}$) needs to deviate by at least $5\%$ from the corresponding SM prediction. While this is clearly beyond the reach of the LHC experiments for the ${\mathcal F}_+$, it is comparable to the expected precision for ${\mathcal F}_L$.\cite{AguilarSaavedra:2007rs}

\section{Conclusions}

Polarization observables in $t\to bW$ decay as represented by the $W$ helicity fractions $\mathcal F_i$ can probe the structure of the $tWb$ vertex and are thus sensitive probes of possible new contributions to top quark interactions beyond the SM.  Such effects can be analyzed using effective theory methods in terms of contributions of higher dimensional effective operators. Within the paradigm of MFV they can also be correlated with other observables, sensitive to new flavor violating contributions, in particular FCNC processes in the down sector.  Then, indirect bounds from $B\to X_s \gamma$ disfavor significant deviations in the $\mathcal F_+$  helicity fraction for individual contributions of dimension-six effective operators, even after taking into account possible significant enhancements due to QCD corrections. On the other hand, the current measurements of $\mathcal F_L$ are already competitive with $B$ physics observables in constraining the effective $tWb$ dipole interactions.

Anomalous $tWd_j$ interactions can also affect $B_{s,d} - \bar B_{s,d}$ mixing phenomenology at one loop. The associated CP violating observables are particularly interesting to consider in light of recently reported anomalies in both $B_{s,d}$ sectors. Within MFV and up to $\mathcal O(m_s/m_b)$ suppressed effects, contributions induced via new $tWd_j$ interactions to $B_{s,d}$ mixing amplitudes are universal. Upon single insertions of individual dimension-six effective operators contributing to $tWb$ interactions, they yield constraints comparable in some cases to $B\to X_s \gamma$ and current direct measurements of $\mathcal F_i$.  On the other hand, taking into account possible large bottom Yukawa effects, several of the MFV allowed effective operators can accommodate the CP violating anomalies and be consistent with constraints from $B\to X_s \gamma$ decay rate measurements. Unfortunately among these possibilities, only one operator predicts observable effects in $t\to bW$ decay. In particular, at least one of the two independent $W$ helicity fractions ${\cal F}_{L,+}$ needs to deviate by at least $5\%$ if this (dipole) operator is solely responsible for the new CP violating effects in $B_{s,d}$ oscillations.  In the future, such CP violating contributions might nonetheless be probed more directly in decays of polarized top quarks, where it is possible to define sensitive CP violating helicity observables.\cite{AguilarSaavedra:2010nx} Such effects could possibly be measured in single top production at the LHC.

\section*{Acknowledgments}
The author would like to thank the organizers of Moriond EW 2011 for the invitation to this exciting conference, as well as Jure Drobnak and Svjetlana Fajfer for a fruitful collaboration on the subject. This work is supported in part by the Slovenian Research Agency.

\end{document}